\documentclass[onecollarge]{svjour2}
\smartqed
\usepackage{graphicx,amssymb,amsmath}

\usepackage{hyperref}
\usepackage{breakurl}
\usepackage{color}
\usepackage{lmodern}
\usepackage{subfig}

\newcommand{\eg}{\textit{e.g.}}  

\newcommand{\ie}{\textit{i.e.}}

\newcommand{\be}{\begin {equation}}
\newcommand{\ee}{\end{equation}}
\newcommand{\bi}{\begin{itemize}}
\newcommand{\ei}{\end{itemize}}
\newcommand{\bea}{\begin {eqnarray}}
\newcommand{\eea}{\end{eqnarray}}
\newcommand{\braket}[2]{\bra{#1}\,#2\rangle} 
\newcommand{\bra}[1]{\langle\,#1\,|}          
\newcommand{\ket}[1]{|\,#1\,\rangle}          

\journalname{Few-Body Systems}
\begin{document}

\title{Advances in Basis Light-front Quantization}

\author{Xingbo~Zhao}

\institute{X.~Zhao\at
Department of Physics and Astronomy, Iowa State University, Ames, IA 50011, USA, \email{xbzhao@iastate.edu}         
}

\date{\today}
\maketitle 
\PACS{11.10.Ef, 11.15.Tk, 12.20.Ds}
\begin{abstract}
        
Basis Light-front Quantization has been developed as a first-principles nonperturbative approach to quantum field theory. In this article we report our recent progress on the applications to the single electron and the positronium system in QED. We focus on the renormalization procedure in this method.

\end{abstract}

\section{Introduction}
Many fundamental questions regarding nuclear and hadron structures and dynamics originate from the nonperturbative aspects of Quantum Chromodynamics (QCD). The Basis Light-front Quantization (BLFQ)~\cite{Vary:2009gt,Honkanen:2010rc} was constructed with the goal of providing a first-principles nonperturbative solution to QCD. So far the development of BLFQ has been carried out in two branches, focusing on the time-dependent and time-independent applications, respectively. For the former, we recently constructed the ``time-dependent'' BLFQ (tBLFQ)~\cite{Zhao:2013cma,Zhao:2013jia}. This method aims to simulate and study the dynamics of quantum field evolution, such as the scattering processes with or without a time-dependent background field. In this article, we, however, limit our focus to the time-independent regime and specifically discuss the application of BLFQ to Quantum Electrodynamics (QED).

This article is organized as follows: in Sec.~\ref{sec:formalism} we explain the basic formalism of BLFQ; in Sec.~\ref{sec:singlee} we review its application to the single electron system in QED; in Sec.~\ref{sec:positronium} we discuss the application to the positronium system and present preliminary numerical results. Finally we conclude in Sec.~\ref{sec:conclusion}.

\section{General Formalism}
\label{sec:formalism}
In this section we briefly review the main aspects of BLFQ, see Refs.~\cite{Vary:2009gt,Zhao:2013jia} for more details.

Basis Light-front Quantization (BLFQ), as a nonperturbative approach to quantum field theory, adopts the light-front dynamics~\cite{Brodsky:1997de} and Hamiltonian formalism. It solves the quantum field system through the eigenvalue problem of the associated light-front Hamiltonian, $P^-$, as,
\begin{align}
    P^-\ket{\beta}=P^-_\beta\ket{\beta}.
    \label{eq:schrodinger}
\end{align}
Upon solving Eq.~(\ref{eq:schrodinger}), one obtains the invariant mass spectrum of the quantum field system through the eigenvalues $P^-_\beta$. The associated light-front amplitudes, $\ket{\beta}$, can be employed to evaluate observables characterizing the structure of these mass eigenstates. 

In order to cast Eq.~(\ref{eq:schrodinger}) into matrix form and solve it numerically, we need to construct a basis for the quantum field system. In BLFQ, we construct the basis in terms of the Fock-space expansion. For each Fock-state particle, we employ a 2D harmonic oscillator (HO) basis to represent its transverse degrees of freedom and a plane-wave basis in the longitudinal direction.  The 2D-HO basis contains a scale parameter $b$, corresponding to the classical amplitude of the ground state HO. The basis functions adopted in BLFQ are closely related to the eigensolutions of the phenomenologically successful light-front holography approach to QCD~\cite{deTeramond:2008ht}.

Next we truncate the basis to make the numerical calculations practical. Basis truncation is performed both on the Fock-sector level and inside each Fock sector. Inside each Fock sector the truncations are implemented independently for the transverse and longitudinal degrees of freedom. 

In the transverse directions we tally the total 2D-HO quantum number in each basis state by $N=\sum_i 2n_i+|m_i|+1$, over all the Fock particles in the basis state. The $n_i$($m_i$) is the radial (angular) quantum number for particle $i$. Then we truncate those whose $N$ exceeds a chosen upper limit $N_{\text{max}}$. 

In the longitudinal direction we compactify the system in a box by imposing the (anti)periodic boundary condition for (fermions) bosons. As a result, the longitudinal momentum of (fermions) bosons can take only positive (half-)integer values (in the unit of inverse box length) and we neglect the zero mode for the bosons. Taking into account of the fact that the total longitudinal momentum of the system, denoted as $K$, is a conserved quantity, $K$ effectively works as a truncation parameter. In BLFQ it is customary to use $K$ as a dimensionless number, representing the longitudinal momentum in the unit of inverse box length. Larger $N_{\text{max}}$ and $K$ result in larger bases, which not only have greater coverage on both the ultraviolet and infrared ends, but also are able to represent finer details of the represented quantum field systems.

These truncations typically break the gauge symmetry in a gauge theory, which leads to difficulties in renormalization. In fact these difficulties have long been known as a standing issue in nonperturbative Hamiltonian-based approaches with Fock-sector truncation, see, \eg, Ref.~\cite{Brodsky:2004cx}. Therefore, one of the central tasks in BLFQ is to explore prospective renormalization procedures which render converging and meaningful results for the observables.

Although the ultimate goal of BLFQ is to address bound state problems in QCD, this method can also be applied to other theories. In order to advance the techniques, we apply BLFQ to QED as an initial step. Below we report our recent work on the single electron and the positronium system in sequence.

\section{Single Electron}
\label{sec:singlee}

The simplest system in QED is a single physical electron. We begin by solving this system in a truncated basis with only $\ket{e}$ and $\ket{e\gamma}$ Fock sectors. 

After factorizing out the center-of-mass motion, one can show that in the $\ket{e}$ sector the light-front amplitude of the
physical electron state involves only one basis state, with both the radial, $n$, and angular quantum number, $m$, being zero.
Due to the quantum fluctuation to the states in the $\ket{e\gamma}$ sector, this state receives self-energy
corrections. According to the sector-dependent renormalization procedure~\cite{Karmanov:2008br,Karmanov:2012aj}, we apply a mass
  counterterm $\Delta m_e$ to this state and only to this state.  Specifically, we replace the mass term $m^2_e/k^+$ from the
  kinetic energy term by $(m_e+\Delta m_e)^2/k^+$, where $k^+$ is the longitudinal momentum of this state. We then evaluate the
  invariant mass $M_e$ of the resulting ground state (identified as the physical electron state $\ket{e_p}$) through
  $M^2_e=\bra{e_p}P^+P^--(P^\perp)^2\ket{e_p}$, with $P^+(P^\perp)$ being the total longitudinal (transverse) momentum
  operator. Next we iteratively adjust the value of $\Delta m_e$ until the resulting invariant mass matches the physical electron
  mass, \ie, $M_e$=0.511\,MeV.

If we calculate observables from the resulting light-front amplitudes, the mass renormalization alone turns out to be insufficient. For example, we encountered problems when we tried to evaluate the anomalous magnetic moment~\cite{Zhao:2014xaa}. We found that the naive result vanishes in the limit of $N_{\max}$ and $K$ approaching infinity, which suggests that the norm of the naive amplitude from diagonalization is incorrect. We attributed this artifact to the violation the Ward-identity caused by Fock-sector truncation.


We solved this problem by proposing an {\it ad-hoc} rescaling~\cite{Zhao:2014xaa} on the direct (naive) amplitude from diagonalization (DA) and then evaluating observables using the rescaled amplitude (RA). The RA is in general 
subdivided into a ``Positive(P)-component'' and an ``Negative(N)-component''. Here ``Positive'' and ``Negative'' are named after 
the sign in front of each component when they appear in Eq.~(\ref{eq:res_norm}) below. Let us denote DA in the single electron problem as $\ket{e_{p}}_{\text D}$, the P-component of RA as $\ket{e_{p}}_{\text P}$ and the N-component as $\ket{e_{p}}_{\text N}$.

\begin{figure}[ht]
\centering
\subfloat[]{
\includegraphics[scale=0.8]{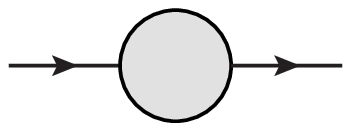}
\label{fig:da_e_dressed}
}
\hspace{3cm}
\subfloat[]{
\includegraphics[scale=0.8]{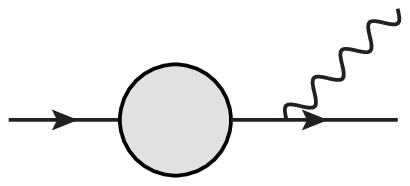}
\label{fig:da_egamma_dressed}
}

\subfloat[]{
\includegraphics[scale=0.8]{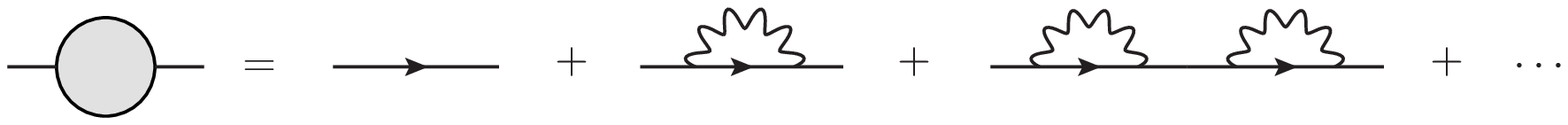}
\label{fig:blob}
}
\caption{Diagrammatic representation of the physical electron amplitude from direct diagonalization (DA). Solid (wavy) line
  represents bare electron (photon). (a): DA in the $\ket{e}$
  sector; (b): DA in the $\ket{e\gamma}$ sector; (c): the blob denotes an infinite sum of repeated self-energy
  corrections (up to the $\ket{e\gamma}$ sector).}
\label{fig:da}
\end{figure}

\begin{figure}[ht]
\centering
\subfloat[]{
\includegraphics[scale=0.8]{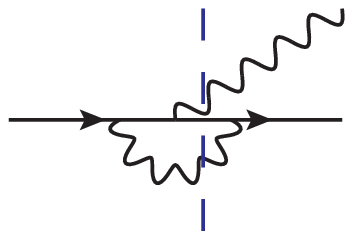}
\label{fig:truncated_vertex}
}
\hspace{3cm}
\subfloat[]{
\includegraphics[scale=0.6]{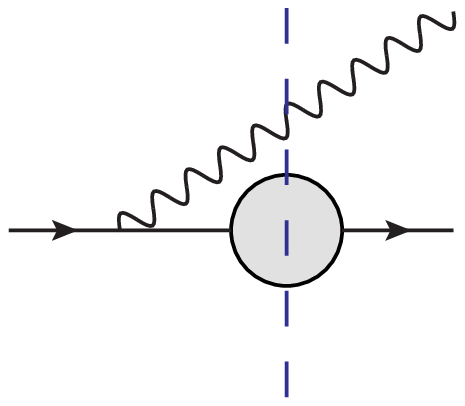}
\label{fig:truncated_self}
}
\caption{Diagrams necessary for satisfying the Ward-identity ($Z_1=Z_2$) but missing in our current truncation level
  ($\ket{e}+\ket{e\gamma}$). (a): vertex corrections; (b): external leg corrections (after photon-emission).}
\label{fig:truncated}
\end{figure}

\begin{figure}[ht]
\centering
\subfloat[]{
\includegraphics[scale=0.8]{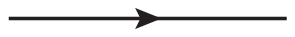}
\label{fig:e_bare}
}
\hspace{1cm}
\subfloat[]{
\includegraphics[scale=0.8]{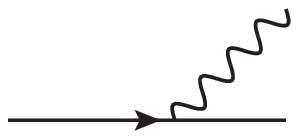}
\label{fig:egamma_bare}
}
\hspace{1cm}
\subfloat[]{
\includegraphics[scale=0.8]{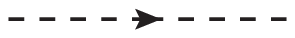}
\label{fig:e_negative_norm}
}
\caption{Diagrammatic representation of the rescaled amplitude (RA). (a): the $\ket{e}$ sector (P-component); (b): the
  $\ket{e\gamma}$ sector (P-component);
  (c): the $\ket{e}$ sector (N-component).}
\label{fig:rescaling}
\end{figure}

Now we recapitulate the rescaling procedure and its motivation. This rescaling procedure is motivated through a diagrammatic
analysis of DA, which is represented in Fig.~\ref{fig:da}.
In our current truncation level, the DA in the $\ket{e\gamma}$ sector only receives external leg corrections {\it before}
  the photon emission. Neither the vertex corrections (Fig.~\ref{fig:truncated_vertex}) nor the external leg corrections {\it
    after} the photon emission (Fig.~\ref{fig:truncated_self}) exist, since these two would require basis states in the
  $\ket{e\gamma\gamma}$ sector. This mismatch of the diagrams causes the violation of the condition $Z_1$=$Z_2$ as required by the
  Ward identity~\cite{Brodsky:2004cx}, where $Z_1$, the vertex correction factor, remains one whereas $Z_2$, the wavefunction
  renormalization factor, is less than one due to the self-energy corrections (Fig.~\ref{fig:blob}).
Our tentative solution is to restore the condition $Z_1=Z_2$ by dividing out the external leg corrections
from DA, which can be read out from the $\ket{e}$ sector contribution to DA (Fig.~\ref{fig:da_e_dressed}). The resulting rescaled
amplitude (RA) is represented in Fig.~\ref{fig:rescaling}.

Following this idea we first read out the wavefunction renormalization factor, $Z_2=|\braket{e}{e_{p}}_{\text D}|^2$, from the
projection of DA onto the sole contributing basis state in the $\ket{e}$ sector. This quantity carries the interpretation of the
probability of finding a bare electron out of a physical electron. Then we rescale DA by $\sqrt{1/Z_2}$ and obtain the P-component of RA,
\begin{align}
    \ket{e_{p}}_{\text P}=\frac{1}{\sqrt{Z_2}}\ket{e_{p}}_{\text D}.
    \label{eq:res_Rp}
\end{align}

In order to maintain the overall norm of RA being one, we introduce the N-component with its norm being negative. Based on
  the observation from perturbation theory~\cite{Brodsky:2000ii}, we hypothesize that in our current truncation level (with the
  $\ket{e\gamma}$ sector as the highest Fock sector) the N-component is contributed solely by the $\ket{e}$ sector. We then fix
  the norm of the N-component from the requirement that the overall norm of RA is one, namely,
\begin{align}
    \ket{e_{p}}_{\text N}=\sqrt{\frac{1-Z_2}{Z_2}}P_e\ket{e_{p}}_{\text P},
    \label{eq:res_Rn}
\end{align}
where $P_e$ is the projection operator onto the $\ket{e}$ sector. 
It is easy to check that the {\it difference} between the squared norm of the P- and N-components is unity (assuming ${}_{\text D}\braket{e_p}{e_p}_{\text D}=1$),
\begin{align}
    1={}_{\text P}\braket{e_{p}}{e_{p}}_{\text P}-{}_{\text N}\braket{e_{p}}{e_{p}}_{\text N}.
    \label{eq:res_norm}
\end{align}
We use the above line of reasoning to suggest that observables $\langle O\rangle$ can be evaluated by sandwiching the pertinent operator $\hat{O}$ with RA according to,
\begin{align}
    \langle O\rangle={}_{\text P}\bra{e_{p}}\hat{O}\ket{e_{p}}_{\text P}-{}_{\text N}\bra{e_{p}}\hat{O}\ket{e_{p}}_{\text N}.
    \label{eq:eval_observ}
\end{align}

Based on Eq.~(\ref{eq:eval_observ}) we evaluated the electron anomalous magnetic moment~\cite{Zhao:2014xaa} and the Generalized
Parton Distribution functions (GPDs) for the electron~\cite{Chakrabarti:2014cwa}. Our numerical results are comparable with those
from perturbation theory at the expected precision.  The agreement on GPDs holds at various $0<x<1$ (the longitudinal
  momentum fraction of the constituent bare electron) and for different momentum transfer $t$, which suggests that the
  renormalization prescription of Eq.~(\ref{eq:eval_observ}) is consistent with perturbative renormalization at a rather
  differential level. We will later apply Eq.~(\ref{eq:eval_observ}) to the Hamiltonian when we evaluate the positronium energy
  spectrum.

\section{Positronium}
\label{sec:positronium}

Now we proceed to the positronium system, a bound state system formed by an electron $e$ and a positron $\bar{e}$. In the previous work we have solved this system by using an effective interaction acting only in the $\ket{e\bar{e}}$ sector~\cite{Wiecki:2014ola,Wiecki:2015xxa}. Now we try to solve this system in QED in a basis consisting of the $\ket{e\bar{e}}$ and $\ket{e\bar{e}\gamma}$ two Fock sectors. Following the ``gauge cutoff'' procedure~\cite{Tang:1991rc}, we include the instantaneous photon interactions only in the $\ket{e\bar{e}}$ Fock sector. We ignore the instantaneous fermion interactions for simplicity.


Since the positronium is not an elementary particle, we expect that the required normalization factors should be extracted from
the single electron system. In contrast to the single electron system, the light-front amplitude of the positronium receives
contributions from multiple basis states in the leading ($\ket{e\bar{e}}$) Fock sector. Each of them is able to receive
self-energy corrections through coupling to the $\ket{e\bar{e}\gamma}$ Fock sector. The basis states in the
  $\ket{e\bar{e}\gamma}$ sector are truncated at a fixed $N_{\text{max}}$ \footnote{In the positronium problem, we elect to use
    independent $N_{\text{max}}$ for different Fock sectors, namely $N_{\text{max1}}$ for the $\ket{e\bar{e}}$ sector and
    $N_{\text{max2}}$ for the $\ket{e\bar{e}\gamma}$ sector, see below. Within each Fock sector the basis states are truncated at
    a uniform $N_{\text{max}}$ (and $K$).} and $K$, whereas each basis state in the $\ket{e\bar{e}}$ sector carries distinct
(longitudinal or transverse) quantum numbers. Hence the ``phase space", or, the ultraviolet and infrared cutoffs, for the
self-energy corrections is different for each individual basis state in the $\ket{e\bar{e}}$ sector. As a result, each basis state
in the $\ket{e\bar{e}}$ sector receives a different self-energy correction. Taking this into account we propose a {\it
  basis-state} dependent renormalization procedure, where distinct renormalization factors ($\Delta m$ and $Z_2$) are applied to
each individual basis state.


In order to determine $\Delta m$ and $Z_2$ (basis) state by (basis) state, we construct a distinct ``embedded single electron" (ESE) system for each basis state in the $\ket{e\bar{e}}$ sector. Similar to the positronium system, the ESE systems span two Fock sectors: $\ket{e\bar{e}}$ and $\ket{e\bar{e}\gamma}$. In the $\ket{e\bar{e}}$ sector we place only one distinct basis state. This basis state consists of one dynamic particle, \eg, the electron, and one spectator, \eg, the positron. The photon is allowed to be coupled only to the dynamic particle, but not the spectator. 
The role of the spectator is to make the kinematics of the self-energy correction for the dynamic particle match that in the positronium system, so that the inferred renormalization quantities can subsequently be used in the fully interacting positronium system.

Before proceeding further, let us briefly recapitulate the assumptions behind the basis-state dependent renormalization
  procedure and the ESE systems: (i) different basis states may be associated with distinct renormalization factors ($\Delta m$
  and $Z_2$); (ii) the renormalization factors for each specific basis state depend on the phase space (in the higher Fock sectors)
  available for the self-energy corrections, which varies from basis state to basis state due to, \eg, the fact that the phase
  space allowed by the truncation parameters in the higher Fock sectors ($N_{\text{max}}$ and $K$ in the $\ket{e\bar{e}\gamma}$
  sector) could be occupied by the spectator particle. For example, different spectator particles take up different longitudinal
  momenta from the total $K$; (iii) the renormalization factors for each constituent particle do not depend on the physical
  systems in which this particle resides. For example, the same renormalization parameters should be applied to a bare electron
  whether it appears in a free physical electron or as a constituent of a positronium state, as long as the available phase space
  for the self-energy correction is the same in both cases.

By solving these ESE systems, we obtain the $\Delta m$ and $Z_2$ for each basis state and then use them to renormalize the positronium system. We first perform mass renormalization by applying each distinct $\Delta m$ into the corresponding kinetic energy term in the positronium system. Upon diagonalization of the Hamiltonian, we obtain the direct amplitude (DA) for the positronium state $\Psi$, which we denote as $\ket{{\Psi}}_{\text D}$.

Next we perform wavefunction rescaling. We first rescale DA and obtain the P-component of RA, $\ket{{\Psi}}_{\text P}$, in the $\ket{e\bar{e}}$ sector. Specifically, we rescale the projection of DA onto each individual basis state $\ket{\beta}$ in the $\ket{e\bar{e}}$ sector with a distinct rescaling factor $Z^{\beta}_2$,
\begin{align}
\braket{\beta}{{\Psi}}_{\text P}=\frac{\braket{\beta}{{\Psi}}_{\text D}}{\sqrt{Z^{\beta}_{2}}}.
\label{eq:Ps_rescl_eebar}
\end{align}
The $Z^{\beta}_{2}$ relates to the basis-state dependent $Z^{e}_{2}$ (for the electron) and $Z^{\bar{e}}_{2}$ (for the positron) evaluated in the ESE systems according to,
\begin{align}
Z^{\beta}_{2}=\frac{Z^e_{2}Z^{\bar{e}}_{2}}{Z^e_{2}+Z^{\bar{e}}_{2}-Z^e_{2}Z^{\bar{e}}_{2}}.
\label{eq:Ps_rescl_eebar_comp}
\end{align}
The derivation of this relation will be shown in the upcoming paper~\cite{Zhao:2014xxx}.

\begin{figure}[ht]
\centering
\subfloat[]{
\includegraphics[scale=0.6]{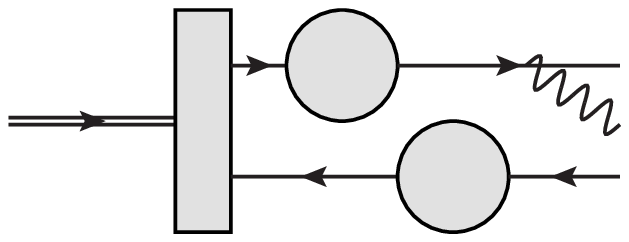}
\label{fig:postronium1}
}
\hspace{3cm}
\subfloat[]{
\includegraphics[scale=0.6]{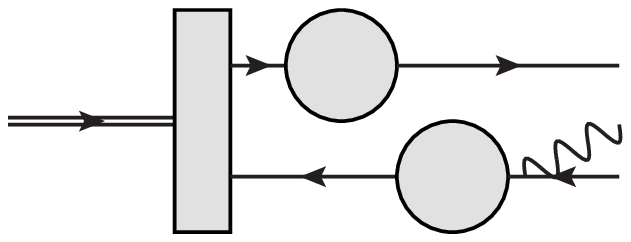}
\label{fig:positronium2}
}
\caption{Diagrammatic representation of the positronium amplitude in the $\ket{e\bar{e}\gamma}$ sector. The shaded rectangle
  represents the positronium amplitude in the $\ket{e\bar{e}}$ sector. The circular blob is the sum of self-energy corrections for the
  constituent electron/positron, as illustrated in Fig.~\ref{fig:blob}.}
\label{fig:positronium_eegamma}
\end{figure}

Now we turn to the $\ket{e\bar{e}\gamma}$ sector. Unlike the electron problem, here the photon can be emitted either from
  $e$ or $\bar{e}$, as illustrated in Fig.~\ref{fig:positronium_eegamma}. 
Therefore it is not straightforward to infer the rescaling
factor from the corresponding ones in the $\ket{e\bar{e}}$ sector as in the case of the electron system. To solve this difficulty, as
an approximation we introduce an ``averaged'' rescaling factor,
\begin{align} 
Z_2^{\text{av}}=\frac{{}_{\text D}\bra{\Psi}P_{e\bar{e}}\ket{\Psi}_{\text D}}{{}_{\text P}\bra{\Psi}P_{e\bar{e}}\ket{\Psi}_{\text P}}, 
\label{eq:Ps_rescl_aver}
\end{align}
where $P_{e\bar{e}}$ is the projection operator onto the $\ket{e\bar{e}}$ sector. We note that $Z_2^{\text{av}}$ is
  weighted by the probability of each basis state appearing in the positronium state $\Psi$. Previous study~\cite{Zhao:2014xaa}
  suggests that $1/Z^e_{2}$ diverges logarithmically as a function of $N_{\text{max}}=K$. If this holds true, in the limit of
  $N_{\text{max}}=K\to\infty$ we expect that nearly all the basis states $\ket{\beta}$ have roughly equal $Z^{\beta}_{2}$ and
  thus the approximation becomes very good\footnote{The basis states with an extremely uneven longitudinal momentum partition
    between $e$ and $\bar{e}$ like \{1/2, K-1/2\}, or with very large (comparable with $N_{\text{max}}$) HO quantum numbers in the
    transverse directions, may still have distinct $Z^{\beta}_{2}$ factors. However, these states are not expected to take
    appreciable probability in well-behaved low-lying positronium states.}. The quality of this approximation will be further
  studied in the upcoming paper~\cite{Zhao:2014xxx}.

With $Z_2^{\text{av}}$ we rescale DA in the $\ket{e\bar{e}\gamma}$ sector as a whole and obtain the P-component of RA in the $\ket{e\bar{e}\gamma}$ sector, 
\begin{align} 
P_{e\bar{e}\gamma}\ket{\Psi}_{\text P}=\frac{P_{e\bar{e}\gamma}\ket{\Psi}_{\text D}}{\sqrt{Z_2^{\text{av}}}}, 
\label{eq:Ps_rescl_eebargamma}
\end{align}
where $P_{e\bar{e}\gamma}$ is the projection operator onto the $\ket{e\bar{e}\gamma}$ sector. Now the norm of $\ket{\Psi}_{\text
  P}$ is $\sqrt{1/Z_2^{\text{av}}}$ (assuming ${}_{\text D}\braket{\Psi}{\Psi}_{\text D}=1$). 

Next, we introduce the N-component of RA, which we take from the $\ket{e\bar{e}}$ sector of the P-component and renormalize it with the norm of $\sqrt{1/Z_2^{\text{av}}-1}$,
\begin{align} 
\ket{\Psi}_{\text N}=\sqrt{\frac{1-Z_2^{\text{av}}}{Z_2^{\text{av}}}}\sqrt{\frac{1}{{}_{\text P}\bra{\Psi}P_{e\bar{e}}\ket{\Psi}_{\text P}}}P_{e\bar{e}}\ket{\Psi}_{\text P}. 
\label{eq:Ps_rescl_neg}
\end{align}
Thus the {\it difference} between the squared norm of the P- and N-components of RA is unity along the lines of Eq.~(\ref{eq:res_norm}), 
\begin{align}
    1={}_{\text P}\braket{\Psi}{\Psi}_{\text P}-{}_{\text N}\braket{\Psi}{\Psi}_{\text N}.
    \label{eq:Ps_res_norm}
\end{align}

Now we are in the position to evaluate observables, $O$, according to, 
\begin{align}
    \langle O\rangle={}_{\text P}\bra{\Psi}\hat{O}\ket{\Psi}_{\text P}-{}_{\text N}\bra{\Psi}\hat{O}\ket{\Psi}_{\text N}.
    \label{eq:Ps_eval_observ}
\end{align}
In this work we focus on the binding energy, $E_b$, of the positronium ground state, which can be evaluated from the light-front energy $P^-_\Psi$ of the system. In order to calculate $P^-_\Psi$ we sandwich the Hamiltonian operator $P^-_{\text{QED}}$ with $\ket{\Psi_{P}}$ and $\ket{\Psi_{N}}$ and obtain the corresponding light-front energy $P^-_\Psi$,
\begin{align}
    \langle P^-_\Psi\rangle={}_{\text P}\bra{\Psi}P^-_{\text{QED}}\ket{\Psi}_{\text P}-{}_{\text N}\bra{\Psi}\bar{P}^{-}_{\text{QED}}\ket{\Psi}_{\text N}.
    \label{eq:Ps_Eb}
\end{align}
Since the N-component of RA contains only the $\ket{e\bar{e}}$ sector, to be consistent with the ``gauge cutoff" procedure, we exclude the instantaneous interaction from the Hamiltonian sandwiched by the N-component, which is denoted by the bar on $\bar{P}^{-}_{\text {QED}}$.

Now we are ready to present the preliminary numerical results for $E_b$. All the results here are calculated at $\alpha=1/\pi$ and with the 2D-HO scale parameter $b$ set to $\alpha M_e$=0.163\,MeV. In the transverse directions we truncate the basis in the $\ket{e\bar{e}}$ and $\ket{e\bar{e}\gamma}$ sectors at $N_{\max1}$ and $N_{\max2}$, respectively. All our calculations are performed at $N_{\max1}$=2, that is, in the $\ket{e\bar{e}}$ sector both the electron and the positron are transversely unexcited, with the 2D-HO quantum numbers of $n=0$ and $m=0$.

\begin{figure}[t]
\centering
\includegraphics[width=0.68\textwidth]{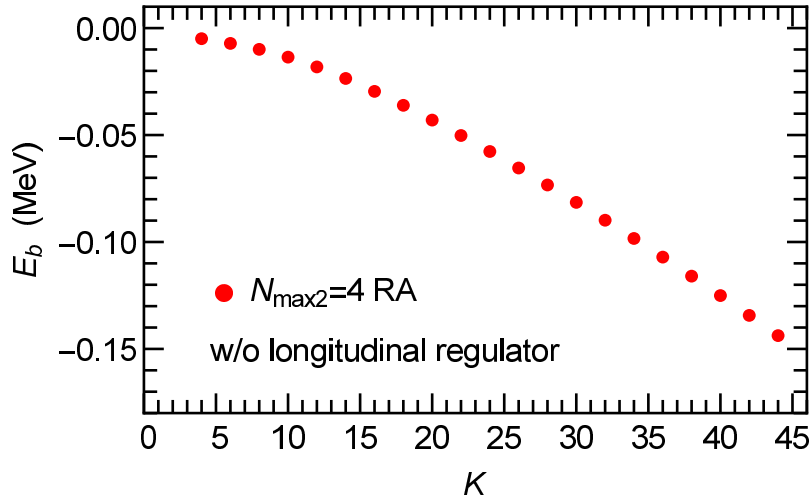}
\caption{\label{fig:Ps_Eb_K_noxregu} (Color online) The ground state binding energy $E_b$ of the positronium as a function of the total longitudinal momentum $K$ {\it before} the longitudinal regulators are applied, (see the text).}
\end{figure}

We first check the longitudinal momentum $K$ dependence of the ground state binding energy in Fig.~\ref{fig:Ps_Eb_K_noxregu}. The results seem to diverge as $K$ increases, which we postulate to be related to the artifacts from the violation of gauge symmetry caused by our basis truncation. In order to regulate this divergence we multiply the following regulator~(\ref{eq:xcutoff_explt}) to the vertex interaction in the Hamiltonian,
\begin{align}
    f_{v}(x)=1-\exp{(-x^2/x^2_c)},
    \label{eq:xcutoff_explt}
\end{align}
and the one in Eq.~(\ref{eq:xcutoff_inst}) to the instantaneous interaction (since the instantaneous interaction is proportional to the electron charge squared $e^2$),
\begin{align}
    f_{i}(x)=[f_{v}(x)]^2=[1-\exp{(-x^2/x^2_c)}]^2.
    \label{eq:xcutoff_inst}
\end{align}
Here $x$ is the longitudinal momentum fraction of the explicit or instantaneous photon and $x_c$ is a cutoff parameter. For consistency we also apply the regulator for the vertex interaction~(\ref{eq:xcutoff_explt}) to the ESE systems.

\begin{figure}[t]
\centering
\hspace{-0.3in}
\includegraphics[width=0.68\textwidth]{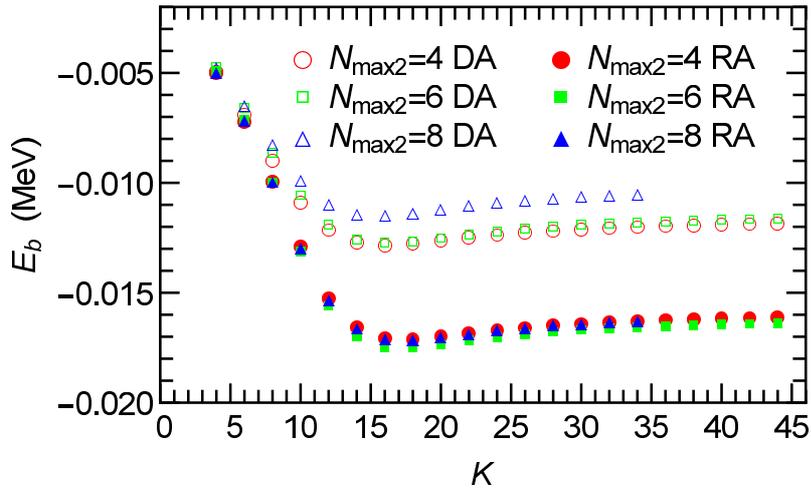}
\caption{\label{fig:Ps_Eb} (Color online) The ground state binding energy $E_b$ of the positronium as a function of the total longitudinal momentum $K$. At $N_{\text{max2}}$=8 we opted for calculating fewer data points to save on the computational resources. Open symbols: $E_b$ evaluated with the light-front amplitudes from direct diagonalization (DA). Solid symbols: $E_b$ evaluated with the rescaled light-front amplitudes (RA). The longitudinal regulators are applied with $x_c=0.052$, (see the text).}
\end{figure}

With these regulators applied, the ground state binding energy converges with $K$. However, the converged value now depends on the
value of $x_c$. In Fig.~\ref{fig:Ps_Eb}, we use $x_c=0.052$, which leads to a converged value around 0.016\,MeV, not far from the
nonrelativistic quantum mechanics value of $E_b$, which is at $\alpha^2m_e/4\sim$0.013\,MeV. Note that our focus here is
  to demonstrate that by a suitable choice of the regulator parameter, a converging result in the vicinity of the expected value can be
  achieved. As the next step we will study the functional dependence of $E_b$ on $x_c$ and try to find independent criteria on
  determining $x_c$ so that the prediction power on the positronium energy spectrum can be maintained.
In Fig.~\ref{fig:Ps_Eb} we also compare the ground state binding energy $E_b$ evaluated with RA from Eq.~(\ref{eq:Ps_Eb}), to that with DA according to $\langle P^-_\Psi\rangle={}_{\text D}\bra{\Psi}P^-_{\text{QED}}\ket{\Psi}_{\text D}$. The convergence with respect to $N_{\text{max2}}$ is improved when RA is used.

\begin{figure}[t]
\centering
\includegraphics[width=0.68\textwidth]{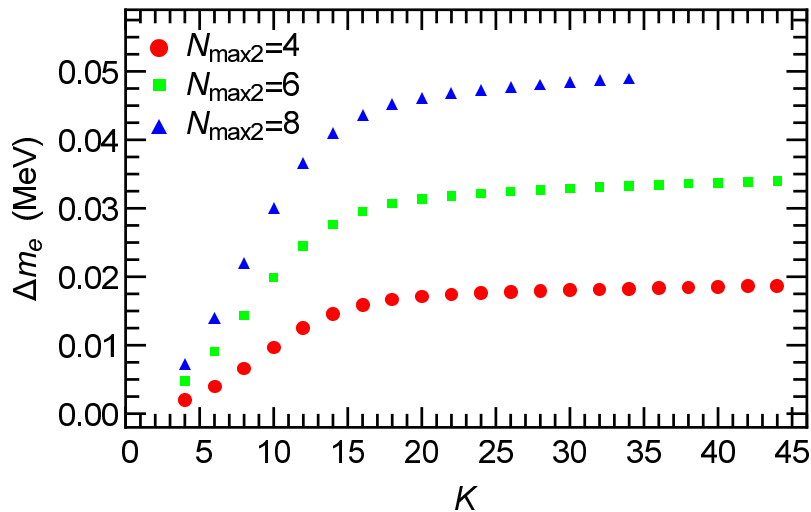}
\caption{\label{fig:delta_me} (Color online) The longitudinal momentum $K$ dependence of the mass counterterm $\Delta m_e$ for the
  basis states in the $\ket{e\bar{e}}$ sector with $n=m=0$ for both $e$ and $\bar{e}$ and with an equal longitudinal momentum partition ($K/2$) between them. The longitudinal cutoff parameter $x_c$ is 0.052. }
\end{figure}

\begin{figure}[t]
\centering
\includegraphics[width=0.68\textwidth]{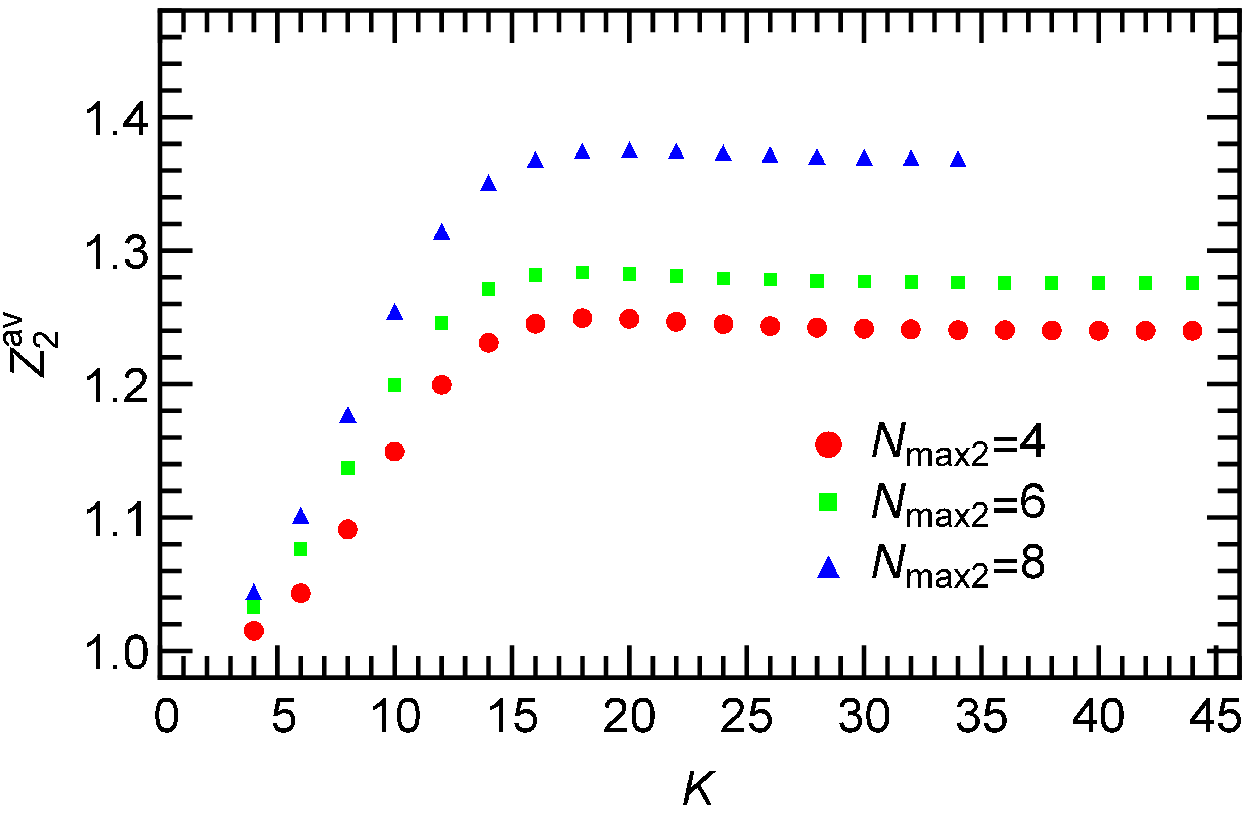}
\caption{\label{fig:Z_2} (Color online) The longitudinal momentum $K$ dependence of the ``averaged" rescaling factor $Z_2^{\text{av}}$ for the ground state positronium. The longitudinal cutoff parameter $x_c$ is 0.052.}
\end{figure}

In Figs.~\ref{fig:delta_me} and~\ref{fig:Z_2} we present the pertinent renormalization quantities which are calculated in the ESE
systems and feed into the positronium system. Fig.~\ref{fig:delta_me} shows the longitudinal momentum $K$ dependence of the mass
counterterm $\Delta m$ for the basis states in the $\ket{e\bar{e}}$ sector with $n=m=0$ for both $e$ and $\bar{e}$ and with an equal longitudinal momentum partition ($K/2$) between them. Fig.~\ref{fig:Z_2} shows the $K$ dependence of the ``averaged" wavefunction renormalization factor $Z_2^{\text{av}}$. 
As $N_{\text{max2}}$ increases, both $\Delta m$ and $Z_2^{\text{av}}$ grow as a result of increasing self-energy correction, whereas the ground state binding energy of the positronium system stays on the same level, cf. Fig.~\ref{fig:Ps_Eb}, which lends support to our renormalization procedure.

We are currently investigating the remaining issues and carrying out further numerical calculations in the larger bases. The updated results will be reported in the forthcoming paper~\cite{Zhao:2014xxx}.



\section{Conclusion and Outlook}
\label{sec:conclusion}
In this article we summarize and report our recent progress on the development of Basis Light-front Quantization (BLFQ), a nonperturbative approach to quantum field theory. We have applied BLFQ to the single electron in QED and the numerical results for its anomalous magnetic moment and generalized parton distributions agree with those from perturbation theory at the expected precision. 

Currently we are extending the application to the positronium system. In order to deal with the renormalization issue, we have proposed a {\it basis-state} dependent renormalization procedure, where we evaluate the pertinent renormalization factors by solving a series of corresponding single electron systems. Each individual system is used to evaluate the renormalization factors for one specific basis state in the positronium system. 

Our next step is to compute the positronium spectrum and the observables, such as the form factors and the (generalized) parton distribution functions, for the low-lying states. Ultimately, our goal is to apply this method to QCD and compute the properties of the bound state systems, including the mesons and baryons.

We acknowledge valuable discussions with K. Tuchin, H. Honkanen, S. J. Brodsky, P. Hoyer, P. Wiecki, Y. Li, P. Maris and J. P. Vary. This work was supported in part by the Department of Energy under Grant Nos. DE-FG02-87ER40371 and DESC0008485 (SciDAC-3/NUCLEI) and by the National Science Foundation under Grant No. PHY-0904782.



\begin{thebibliography}{99}

%
%
\bibitem{Vary:2009gt}
  J.~P.~Vary, H.~Honkanen, J.~Li, P.~Maris, S.~J.~Brodsky, A.~Harindranath, G.~F.~de Teramond, P.~Sternberg,  E.G.~Ng and C.~Yang,
  Phys.\ Rev.\ C {\bf 81} (2010) 035205.
  
%

\bibitem{Honkanen:2010rc}
  H.~Honkanen, P.~Maris, J.~P.~Vary and S.~J.~Brodsky,
  Phys.\ Rev.\ Lett.\  {\bf 106} (2011) 061603.
%
%
%

\bibitem{Zhao:2013cma}
  X.~Zhao, A.~Ilderton, P.~Maris and J.~P.~Vary,
  Phys.\  Rev.\  D {\bf 88} (2013) 065014.
%
\bibitem{Zhao:2013jia}
  X.~Zhao, A.~Ilderton, P.~Maris and J.~P.~Vary,
  Phys.\  Lett.\  B {\bf 726} (2013) 856.
 
\bibitem{Brodsky:1997de}
  S.~J.~Brodsky, H.-C.~Pauli and S.~S.~Pinsky,
  Phys.\ Rept.\  {\bf 301} (1998) 299.
  
\bibitem{deTeramond:2008ht}
  G.~F.~de Teramond and S.~J.~Brodsky,
  Phys.\ Rev.\ Lett.\  {\bf 102} (2009) 081601.


\bibitem{Brodsky:2004cx}
  S.~J.~Brodsky, V.~A.~Franke, J.~R.~Hiller, G.~McCartor, S.~A.~Paston and E.~V.~Prokhvatilov,
  Nucl.\ Phys.\ B {\bf 703} (2004) 333.

  
 \bibitem{Karmanov:2008br}
   V.~A.~Karmanov, J.-F.~Mathiot and A.~V.~Smirnov,
 Phys.\ Rev.\ D {\bf 77} (2008) 085028.

\bibitem{Karmanov:2012aj}
  V.~A.~Karmanov, J.~F.~Mathiot and A.~V.~Smirnov,
  Phys.\ Rev.\ D {\bf 86} (2012) 085006.

\bibitem{Zhao:2014xaa}
  X.~Zhao, H.~Honkanen, P.~Maris, J.~P.~Vary and S.~J.~Brodsky,
  Phys.\ Lett.\ B {\bf 737} (2014) 65.


\bibitem{Brodsky:2000ii}
  S.~J.~Brodsky, D.~S.~Hwang, B.~Q.~Ma and I.~Schmidt,
  Nucl.\ Phys.\ B {\bf 593} (2001) 311.


\bibitem{Chakrabarti:2014cwa}
  D.~Chakrabarti, X.~Zhao, H.~Honkanen, R.~Manohar, P.~Maris and J.~P.~Vary,
  Phys.\ Rev.\ D {\bf 89} (2014) 11,  116004.
  
  

\bibitem{Wiecki:2015xxa}
  P.~Wiecki, Y.~Li, X.~Zhao, P.~Maris and J.~P.~Vary,
  Few Body Syst.\  (2015),
  arXiv:1502.02993 [nucl-th].

\bibitem{Wiecki:2014ola}
  P.~Wiecki, Y.~Li, X.~Zhao, P.~Maris and J.~P.~Vary,
  Phys.\ Rev.\ D {\bf 91} (2015) 10,  105009.

 
\bibitem{Tang:1991rc}
  A.~C.~Tang, S.~J.~Brodsky and H.~C.~Pauli,
  Phys.\ Rev.\ D {\bf 44} (1991) 1842.

\bibitem{Zhao:2014xxx} 
 X.~Zhao, Y.~Li, P.~Maris and J.~P.~Vary,
 in preparation.
 
 
%
%
%
%
%
%
%
%
%
%
%
%
%

%
%



\end{thebibliography}
\end{document}